\documentclass[preprint,epsfig,floats,aps]{revtex4}
\usepackage{epsfig}
\begin{document}
\topmargin=-1.5cm
\title{Behaviors of the charge fluctuation in \\  
       relativistic nucleus-nucleus collisions}
\author{Ben-Hao Sa$^{1,2,4,5,}$\footnote{Email: sabh@iris.ciae.ac.cn}, 
 Xu Cai$^{4,2}$, Zong-Di Su$^{1,4}$, An Tai$^3$, and Dai-Mei Zhou$^2$}
\affiliation{
$^1$  China Institute of Atomic Energy, P. O. Box 275 (18),
      Beijing, 102413 China \\
$^2$  Institute of Particle Physics, Huazhong Normal University,
      Wuhan, 430079 China\\
$^3$  Department of Physics and Astronomy, University of California,
      at Los Angeles, Los Angeles, CA 90095 USA \\
$^4$  CCAST (World Lab.), P. O. Box 8730 Beijing, 100080 China\\
$^5$  Institute of Theoretical Physics, Academia Sinica, Beijing,
      100080 China
}
\begin{abstract}
Using a hadron and string cascade model, JPCIAE, we investigated the 
dependence of event-by-event charge fluctuation on the energy, centrality, 
window size, resonance decay, and the final state interaction for $Pb+Pb$ 
collisions at SPS and LHC energies and $Au+Au$ collisions at RHIC 
energies. The JPCIAE results of charge fluctuation as a function of the 
rapidity window size in $Pb+Pb$ collisions at SPS energies were compared  
with the preliminary NA49 data. Comparisons with STAR and PHENIX data of 
$Au+Au$ collisions at $\sqrt{s_{nn}}$=130 GeV were also given. It seems that 
the final state interaction and resonance decay play a gentle role on the 
charge fluctuations. The charge fluctuations are slightly decreasing with or 
nearly independent of the reaction energy and hardly depend on the 
collision centrality.
\end{abstract}
\maketitle
 
In \cite{sa0,bon}, the energy fluctuation (heat capacity) was first related to 
the liquid-gas phase transition in intermediate energy heavy-ion collisions.  
The irregular behavior of heat capacity was then proposed to 
study the phase transition from hadronic matter to a Quark-Gluon-Plasma (QGP) 
provided that the event-by-event (E-by-E) fluctuation of temperature is 
observable in relativistic nucleus-nucleus collisions \cite{sto,ste}. Such 
irregular behavior is also a characteristic of a phase transition: a jump in a 
first order phase transition and a singularity in a second order one 
\cite{sto}. The E-by-E fluctuation of an observable might supply important  
information such as the hadronic matter compressibility \cite{mro}, the 
position and property of a critical point in the QCD phase diagram of 
temperature T vs. chemical potential $\mu$ \cite{ste}, etc. .  
In \cite{ste} it was also predicted that the E-by-E fluctuation pattern in 
average transverse momentum, for instance, would significantly be changed 
around a critical point.

\begin{figure}[ht]
\centerline{\hspace{-0.5in}
\epsfig{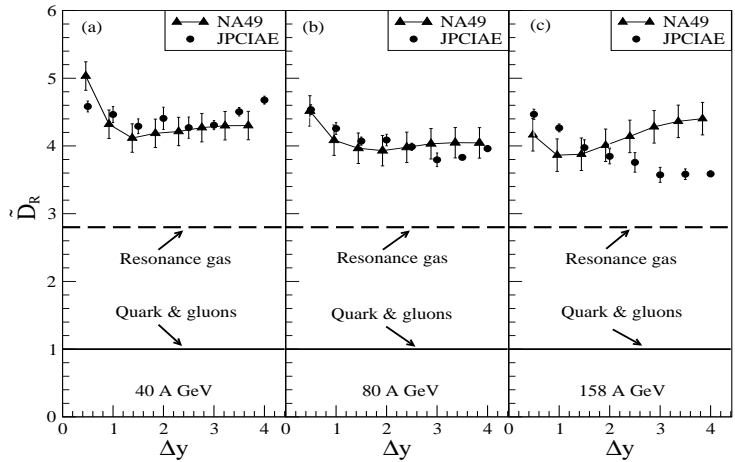}}
\caption{$\tilde{D_R}$ as a function of the $\Delta y$ in 40, 80 
and 158A GeV/c $Pb+Pb$ collisions. The preliminary NA49 data were 
take from \cite{fri}.}
\label{fig1_flu1}
\end{figure}

With the increase of interaction energy a rather high particle multiplicity 
is accessible and the statistically significant measurements of E-by-E 
fluctuation became possible for the first time in $Pb+Pb$ collisions at 158A 
GeV/c \cite{rol,afa0,ble,app,afa} and recently in $Au+Au$ collisions at 
$\sqrt{s_{nn}}$ =130 GeV \cite{phe}. Though it was claimed that the 
non-statistical contributions to E-by-E fluctuation of the average transverse 
momentum, the $k/\pi$ ratio, and the net charge multiplicity are small 
\cite{app,afa,phe}, the calculations of E-by-E fluctuations based on hadronic 
transport models \cite{ble,liu,cap} and effective models \cite{vol,bay,hei} 
were stimulated. 
 
\begin{figure}[ht]
\centerline{\hspace{-0.5in}
\epsfig{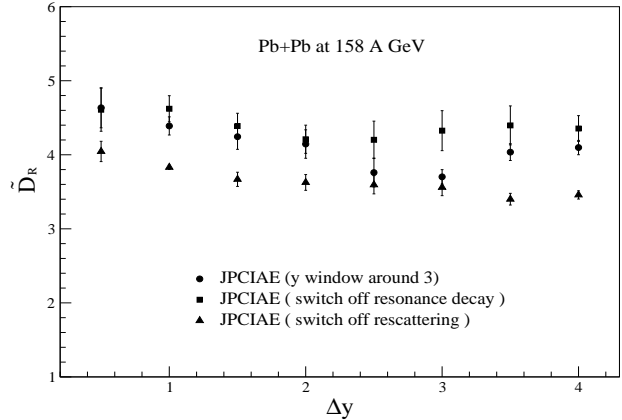}}
\caption{The effects of rescattering and resonance decay on the 
distribution of $\tilde{D_R}$ vs. $\Delta y$ in 158A GeV/c $Pb+Pb$ collisions  
obtained with JPCIAE.} 
\label{fig2_flu1}
\end{figure}

Since the unit of charge (baryon charge) in the QGP phase is 1/3 while it is 
1 in the hadronic phase, the thermal model predicted that the value of the   
charged particle ratio E-by-E fluctuation, $D_R$ (defined below), in the 
hadronic phase would be a factor of $\sim$ 2.5 - 4 larger than that in the 
QGP phase \cite{jeo,asa,lin}. The charged particle ratio E-by-E fluctuation  
was then proposed as a signal of QGP formation if the initial fluctuations 
survive hadronization and their relaxation time is longer than the collision 
time \cite{jeo,asa}. In \cite{jeo1} UrQMD \cite{uqmd} was used to investigate 
the charged particle ratio fluctuation in $Pb+Pb$ collisions at SPS and 
$Au+Au$ collisions at the full RHIC energy. However, the UrQMD predictions 
for the charged particle ratio fluctuation in $Pb+Pb$ collisions at SPS 
energy were around 3 while the preliminary NA49 data \cite{fri} were around 4.
         
The hadron and string cascade model, JPCIAE, was employed in this paper to 
further study the charge fluctuations. The model results were compared with 
the preliminary NA49 data of the charged particle ratio fluctuation as a 
function of the rapidity window size in $Pb+Pb$ collisions at 40, 80 and 
158A GeV/c \cite{fri} and with STAR and PHENIX data in $Au+Au$ 
collisions at $\sqrt{s_{nn}}$=130 GeV \cite{volo,phe}. Meanwhile, the 
dependence of charge fluctuations on reaction energy (from SPS up to LHC), 
centrality (impact parameter b), final state interaction (rescattering), and 
the resonance decay ($\rho$ and $\omega$) was investigated. This study shows 
that the charge fluctuations are slightly decreasing with or nearly 
independent of the reaction energy and hardly depend on the collision 
centrality. The charge fluctuations are gently affected by the rescattering 
and the resonance decay ($\rho$ and $\omega$). 

The JPCIAE model was developed based on PYTHIA \cite{sjo1}, which is a well 
known event generator for hadron-hadron collisions. In the JPCIAE model the 
radial position of a nucleon in colliding nucleus A (indicating the atomic 
number of this nucleus as well) is sampled randomly according to the 
Woods-Saxon distribution and the solid angle of the nucleon is sampled 
uniformly in 4$\pi$. Each nucleon is given a beam momentum in z direction and 
zero initial momenta in x and y directions. The collision time of each 
colliding pair is calculated under the requirement that the least approach 
distance of the colliding pair along their straight line trajectory 
(mean field potential is not taken into account in JPCIAE) should be 
smaller than $\displaystyle{\sqrt{\sigma_{tot}/\pi}}$. Here $\sigma_{tot}$ 
refers to the total cross section. The nucleon-nucleon collision with the 
least collision time is then selected from the initial collision list to 
perform the first collision. Both the particle list and the collision list are 
then updated such that the new collision list may consist of not only nucleon-
nucleon collisions but also collisions between nucleons and produced 
particles and between produced particles themselves. The next collision is 
selected from the new collision list and the processes above are 
repeated until the collision list is empty.

For each executing collision pair, if its CMS energy is above a certain 
threshold (=4 GeV in program), we assume that strings are formed after 
the collision and PYTHIA is used to deal with particle production. Otherwise, 
the collision is treated as a two-body collision \cite{cugn,bert,tai1}. The 
threshold above is chosen in such a way that JPCIAE correctly reproduces the 
charged multiplicity distributions in nucleus-nucleus collisions \cite{sa1}. 
It should be noted here that the JPCIAE model is not a simple superposition of 
nucleon-nucleon collisions since the rescatterings of secondary particles 
are taken into account. We refer to \cite{sa1} for more details about the 
JPCIAE model. Note that particle production from strings in JPCIAE is 
determined by the Lund fragmentation scheme \cite{lund}, in which only the 
lowest excitation state of a resonance is included.

\begin{figure}[ht]
\centerline{\hspace{-0.5in}
\epsfig{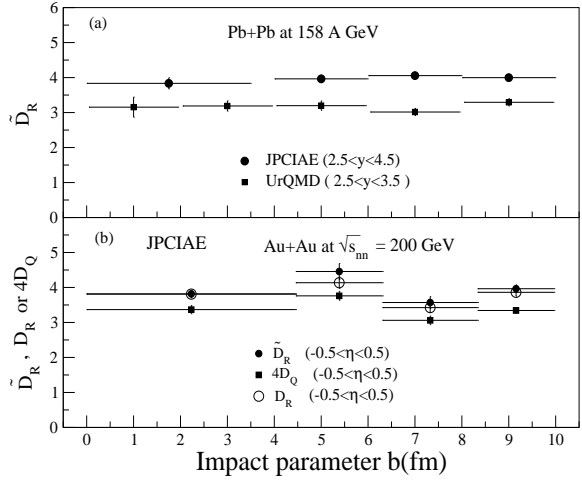}}
\caption{The centrality dependence of charge fluctuations.} 
\label{fig3_flu1}
\end{figure}

If the deviation (i. e. fluctuation \cite{rei}) of a physical variable $x$ 
from its average value per event $<x>$ is defined as
 \begin{equation}
 \delta x=x-<x>,
 \end{equation}  
the variance of $x$ reads \cite{rei}
 \begin{equation}
 <(\delta x)^2>=<x^2>-<x>^2.
 \end{equation} 
Suppose $\displaystyle{x\equiv R=N_+/N_-}$ to be the ratio of positively to 
negatively charged particle multiplicity, the corresponding variance is 
 \begin{equation}
 <(\delta R)^2>=<R^2>-<R>^2.
 \end{equation}
Similarly the variance of net charge multiplicity, $\displaystyle{Q=N_+-N_-}$, 
reads
 \begin{equation}
 <(\delta Q)^2>=<Q^2>-<Q>^2.
 \end{equation}
However, what is interesting is not $\displaystyle{<(\delta R)^2>}$ or  
$\displaystyle{<(\delta Q)^2>}$ but 
 \begin{equation}
 D_R\equiv<N_{ch}><(\delta R)^2>
 \end{equation} 
or 
 \begin{equation}
 D_Q\equiv\frac{<(\delta Q)^2>}{<N_{ch}>},
 \end{equation}
where $\displaystyle{N_{ch}=N_++N_-}$ refers to the total charge multiplicity. 
A relation follows approximately \cite{jeo}
 \begin{equation}
 D_R\simeq4D_Q.
\label{dd}
 \end{equation} 
The thermal (effective) model predictions for $D_R$ are \cite{jeo}: $\sim$4 
for a pion gas, $\sim$3 for a resonance pion gas (pions from $\rho$ and 
$\omega$ decays), and $\sim$0.75 for massless noninteracting quarks and gluons 
(that is $\sim$1 from lattice calculations).   

\begin{figure}[ht]
\centerline{\hspace{-0.5in}
\epsfig{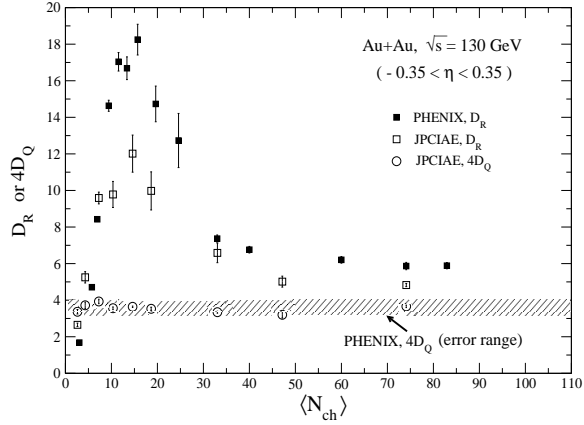}}
\caption{The charge fluctuations as a function of charged multiplicity, 
         $<N_{ch}>$.} 
\label{fig4_flu1}
\end{figure}

As mentioned in \cite{jeo}, one main assumption made in the thermal model 
predictions is that the studied system can be described as a grand canonical 
ensemble. However, in experiments or dynamical simulations, the investigated 
subsystem (e. g. within a rapidity interval $\Delta y$) is a finite fraction 
of the full system (e. g. in the full rapidity region). Therefore, the 
assumption of a grand canonical ensemble is only valid in the limit of 
$\displaystyle{<N_{ch}>_{\Delta y}/<N_{ch}>_{total}}\rightarrow$ 0, such that 
the rest system plays the role of a thermal resource. In order to compare the 
experiments or dynamical simulations with the thermal model predictions it 
might be better introducing a correction factor \cite{jeo}
 \begin{equation}
 C_y=1-\frac{<N_{ch}>_{\Delta y}}{<N_{ch}>_{total}}.
 \end{equation}  
Another assumption adopted in the thermal model predictions is the vanishing 
of net charge \cite{jeo}. However, that is actually impossible in experiments 
or dynamical simulations, the corresponding correction factor \cite{jeo} 
reads 
 \begin{equation}
 C_{\mu}=\frac{<N_+>_{\Delta y}^2}{<N_->_{\Delta y}^2}.
 \end{equation}
The $\displaystyle{D_R}$ with corrections above is denoted as       
 \begin{equation}
 \tilde{D_R}=\frac{D_R}{C_yC_{\mu}}. 
 \end{equation}

The fluctuation is usually composed of statistical fluctuation and dynamical 
fluctuation. There are many sources to be considered as dynamical 
fluctuations, such as string fragmentation (or QCD color fluctuations), 
centrality (impact parameter or participants), rescattering, 
resonance decay, etc. . On the contrary, the statistical fluctuation is no 
dynamical origin and could be described in a stochastic scenario by 
probability distribution functions \cite{phe,rei}. Only a finite number of 
events could be generated in experiments or dynamical simulations causes also 
the statistical fluctuation. Though it is necessary to study the influences of 
reaction energy, centrality, rescattering, and resonance decay individually, 
an alternative way to investigate the non-statistical contribution is to 
compare E-by-E fluctuation distribution extracted from real events with ones 
from mixed events \cite{afa}. The mixed events here are constructed from the 
real events so that in principle only statistical fluctuation survives in the 
mixed events \cite{afa}.     

\begin{figure}[ht]
\centerline{\hspace{-0.5in}
\epsfig{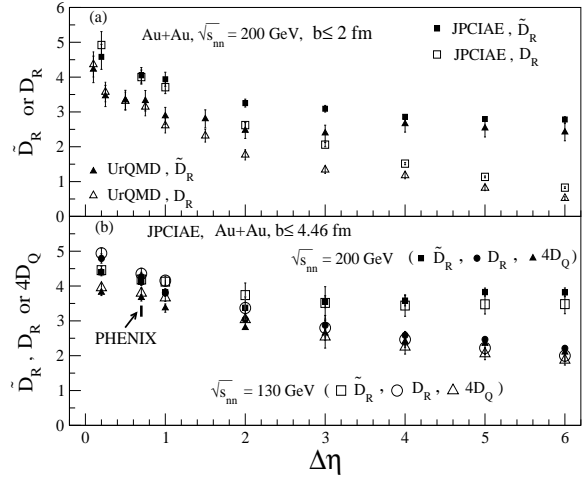}}
\caption{The $\tilde{D_R}$, $D_R$, and 4$D_Q$ as a function of $\Delta \eta$:
(a) comparison of the results from JPCIAE and UrQMD for $Au+Au$ 
collisions at $\sqrt{s_{nn}}$=200 GeV; (b) comparison of the JPCIAE results 
for $Au+Au$ collisions at $\sqrt{s_{nn}}$=130 and 200 GeV.} 
\label{fig5_flu1}
\end{figure}

In Fig. 1 the JPCIAE results of $\tilde{D_R}$ as a function of $\Delta y$ in 
40, 80 and 158A GeV/c $Pb+Pb$ collisions (full circles) are compared to the 
NA49 preliminary data (full triangles) \cite{fri}. Corresponding to the 
centrality cut of 7.2\% at 40 and 80A Gev/c and 10\% at 158A GeV/c in the NA49 
experiments the impact parameters in the JPCIAE calculations were set to be 
b$\leq$3.57 fm and b$\leq$4.20 fm, respectively. $\Delta y$ was set around 
2.9, 3.2, and 3.6, respectively, for 40, 80, and 158A GeV/c energies, and the 
$p_t$ window was set to be $0.005<p_t<2.5$ GeV/c for all the three beam 
momenta as in the NA49 experiments. The dashed and solid lines in this figure 
are the thermal model predictions for a resonance pion gas and the lattice 
Monte Carlo result for a quark-gluon gas, respectively \cite{jeo1}. One sees 
from this figure that the JPCIAE results are generally compatible with the 
preliminary NA49 data for 40 and 80A GeV/c $Pb+Pb$ collisions. However, for 
158A GeV/c $Pb+Pb$ collisions there exists discrepancies in the $\Delta y$ 
dependence between preliminary NA49 data and JPCIAE results. Such differences 
are not due to statistics and require further studies.

The effects of rescattering and resonance decay ($\rho$ and $\omega$ primarily 
\cite{jeo}) on the distribution of $\tilde{D_R}$ vs. $\Delta y$ are shown in 
Fig. 2 for 158A GeV/c $Pb+Pb$ collisions. In this figure, the circles, 
the triangles, and the squares are,  respectively, the results of default 
JPCIAE, JPCIAE without rescattering, and JPCIAE without $\rho$ and $\omega$ 
resonance decays. In the JPCIAE calculations the impact parameter was 
b$\leq$3.5 fm,  $\Delta y$ was set around 3 and the $p_t$ window was $0<p_t<5$ 
GeV/c. Globally speaking, the rescattering effect is gentle, that is 
consistent with the conclusion from the RQMD model \cite{zhang}. The effect of 
resonance decay seems weak, too and the shift in $\tilde{D_R}$ is 
smaller than 1 over all the $\Delta y$ region unlike what is expected based on 
the thermal model predictions for a pion gas and resonance pion gas. However, 
in the default JPCIAE calculations no all the mesons are from $\rho$ and 
$\omega$ resonance decays, which may explain in part why the shift is smaller 
than 1. 

Fig. 3 (a) compares the JPCIAE results (circles) of centrality dependence of 
$\tilde{D_R}$ in $Pb+Pb$ collisions at 158A GeV/c with UrQMD results (squares, 
taken from \cite{jeo1}). In both calculations the rapidity window was 
$2.5<y<4.5$. The discrepancies between JPCIAE and UrQMD results might 
attribute in part to the higher resonance states included in the UrQMD model. 
In Fig. 3 (b) the centrality dependences of $\tilde{D_R}$, $D_R$, and 4$D_Q$ 
in $Au+Au$ collisions at $\sqrt{s_{nn}}$=200 GeV from JPCIAE (-0.5$<\eta<$0.5) 
are given by full circles, open circles, and full squares, respectively. One 
sees from Fig. 3 that the charge fluctuation measures (i. e. the $\tilde{D_R}$
, $D_R$, and 4$D_Q$) are not so sensitive to the impact parameter. That is 
consistent with the STAR and PHENIX corresponding observations 
\cite{volo,phe}. 

\begin{figure}[ht]
\centerline{\hspace{-0.5in}
\epsfig{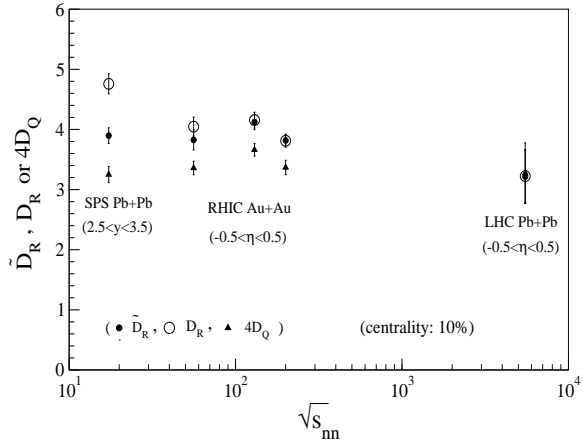}}
\caption{The energy dependence of $\tilde{D_R}$, $D_R$, and 4$D_Q$.} 
\label{fig6_flu1}
\end{figure}

In Fig. 4 the JPCIAE results of $D_R$ and 4$D_Q$ as a function of $<N_{ch}>$ 
in peripheral $Au+Au$ collisions at $\sqrt{s_{nn}}$=130 GeV (-0.35 $<\eta<$ 
0.35, $p_t >$ 0.2) were compared with PHENIX data \cite{phe}. For simplicity 
only part of $D_R$ data points were copied (full squares with error bar) and 
compared with JPCIAE results (open squares with error bar). Most $D_R$ results 
from JPCIAE were lower than PHENIX data, at peak region especially. That 
might attribute in part to the PHENIX spectrometer has an acceptance of 
$\pi/2$ radians in azimuthal angle, since a linear extrapolation to full 
azimuthal coverage leads to the decreasing of fluctuation measures \cite{phe}. 
In Fig. 4 the PHENIX data of 4$D_Q$ with error bar were denoted simply by 
shaded region and compared with JPCIAE results of 4$D_Q$ (open circles). One 
sees that the JPCIAE results of 4$D_Q$ are compatible with PHENIX data.    
      
Fig. 5 (a) compared the JPCIAE results of $\tilde{D_R}$ and $D_R$ as a 
function of $\Delta\eta$ (b$\leq$2 fm) in $Au+Au$ collisions at $\sqrt{s_{nn}}
$=200 GeV with the UrQMD results (b$\leq$2 fm, taken from \cite{jeo1} where 
$\Delta y$ was used). The full and open squares and the full and open 
triangles in this panel are, respectively, the results of $\tilde{D_R}$ and 
$D_R$ from JPCIAE and UrQMD. Generally speaking, the results of JPCIAE are 
systematically higher than those of UrQMD, similar to Fig. 3 (a). In Fig. 5 
(b) the JPCIAE results of $\tilde{D_R}$, $D_R$ and 4$D_Q$ as a function of 
$\Delta\eta$ in $Au+Au$ collisions at $\sqrt{s_{nn}}$=130 (open squares,  
circles, and triangles, respectively) and at 200 GeV (full squares, circles, 
and triangles, respectively) are compared with each other. In those 
calculations the centrality and $p_t$ cuts were, respectively, 10\% 
most central collisions and $p_t>$0.2 . The thick stick at 
$\Delta\eta$=0.7 is the PHENIX datum \cite{phe} of 4$D_Q$ in $Au+Au$ 
collisions at $\sqrt{s_{nn}}$=130 GeV, which is about 10\% lower than the  
corresponding JPCIAE result (open triangle). It is also interesting to note 
that the JPCIAE result of $D_Q\sim$0.9 at $\Delta\eta$=1 in $Au+Au$ collisions 
at $\sqrt{s_{nn}}$=130 GeV is about 10\% higher than the corresponding STAR 
datum 0.8 extracted under the assumption of zero net charge \cite{volo}. From 
Fig. 5 (b) one sees that globally speaking the charge fluctuation measures are 
not sensitive to the change of energy from 130 to 200 GeV, which is consistent 
with the conclusions in \cite{asa,jeo1,zhang}. 

Finally, the JPCIAE results of energy dependence of $\tilde{D_R}$ (full 
circles), $D_R$ (open circles), and 4$D_Q$ (full triangles) from SPS to RHIC 
and then to LHC energy were given in Fig. 6 . In those calculations the 
centrality cuts were all set to be 10\% most central collisions and rapidity 
windows are 2.5$<y<$3.5 for $Pb+Pb$ at SPS and -0.5$<\eta<$0.5 for $Au+Au$ at 
RHIC and $Pb+Pb$ at LHC energy, respectively. One sees from Fig. 6 that the 
$D_R$ might be decreasing slightly with energy. However, the $\tilde{D_R}$ 
and 4$D_Q$ show almost no energy dependence within error bar.
  
In summary, a hadron and string cascade model, JPCIAE, has been employed in 
this paper to investigate the energy, centrality, rescattering, and resonance  
decay dependences of the charge fluctuation measures. Within the framework of 
this model the calculated results seem compatible with the preliminary 
NA49 data for $Pb+Pb$ collisions at 40 and 80A GeV/c. For 158A 
GeV/c $Pb+Pb$ collisions there exists discrepancies in the $\Delta y$ 
dependence between JPCIAE results and preliminary NA49 data. The 
JPCIAE results for $Au+Au$ collisions at $\sqrt{s_{nn}}$=56, 130 and 200 GeV 
and for $Pb+Pb$ collisions at $\sqrt{s_{nn}}$=5500 GeV were given as well. 
Comparisons between JPCIAE results and experimental data from STAR and PHENIX 
were also made for $Au+Au$ collisions at $\sqrt{s_{nn}}$=130 GeV. It seems 
that the charge fluctuation measures are nearly independent of the collision 
centrality. Their dependence on the reaction energy is weak. It is also found 
that the effect of resonance decay ($\rho$ and $\omega$) on the charge 
fluctuation measures is gentle. However, the rescattering effect might be 
somewhat stronger than resonance decay.  

Finally, the financial supports from NSFC (19975075, 10135030, and 10075035) 
in China and DOE in USA are acknowledged.

\end{document}